\pdfoutput=1
\documentclass[a4paper,referee]{raa}         

\usepackage{graphicx,times}
\usepackage{natbib}
\usepackage{amssymb,amsmath}
\usepackage{colortbl} 
\usepackage{xcolor}  
\usepackage{soul}     
\usepackage{tikz}
\bibpunct{(}{)}{;}{a}{}{,}

\usepackage[pagebackref=true]{hyperref}

\begin{document}

   \title{The Mini-SiTian Array: real-bogus classification using deep learning}

 \volnopage{ {\bf 20XX} Vol.\ {\bf X} No. {\bf XX}, 000--000}
   \setcounter{page}{1}

   \author{Jing-Hang Shi
   \inst{1,2,3}, Hong-Rui Gu\inst{1,2}, Yang Huang\inst{1,3}\textsuperscript{*}, Yan-Xia Zhang\inst{1,2}\textsuperscript{*}, Peng-Liang Du\inst{2}
   }

   \institute{CAS Key Laboratory of Optical Astronomy, National Astronomical Observatories, Chinese Academy of Sciences, Beijing 100101, People's Republic of China; \\
   \and National Astronomical Observatories, Chinese Academy of Sciences, Beijing 100101, People’s
Republic of China
   \and School of Astronomy and Space Science, University of Chinese Academy of Sciences, 19A Yuquan Road, Shijingshan District, Beijing 100049, People’s Republic of China\\
   Corresponding author: Yang Huang (huangyang@ucas.ac.cn); Yan-Xia Zhang (zyx@bao.ac.cn)\\
\vs \no
   {\small Received 20XX Month Day; accepted 20XX Month Day}
}

\abstract{The Mini-SiTian (MST) project is a pathfinder for China's next-generation large-scale time-domain survey, SiTian, aimed at discovering variable stars, transients, and explosive events. MST generates hundreds of thousands of transient alerts every night, approximately 99\% of which are false alarms, posing a significant challenge to its scientific goals. To mitigate the impact of false positives, we propose a deep learning–based solution and systematically evaluate thirteen convolutional neural networks. The results show that ResNet achieves exceptional specificity (99.70\%), EfficientNet achieves the highest recall rate (98.68\%), and DenseNet provides balanced performance with a recall rate of 94.55\% and specificity of 98.66\%. Leveraging these complementary strengths, we developed a bagging-based ensemble classifier that integrates ResNet18, DenseNet121, and EfficientNet\_B0 using a soft voting strategy. This classifier achieved the best AUC value (0.9961) among all models, with a recall rate of 95.37\% and specificity of 99.25\%. It has now been successfully deployed in the MST real-time data processing pipeline. Validation using 5,000 practically processed samples with a classification threshold of 0.798 showed that the classifier achieved 88.31\% accuracy, 91.89\% recall rate, and 99.82\% specificity, confirming its effectiveness and robustness under real application conditions.
}

   \authorrunning{J.-H. Shi et al. }            
   \titlerunning{real-bogus classification using deep learning}  
   \maketitle

%
\section{Introduction} \label{sec:intro}

The universe is a dynamic system where the properties of celestial objects, such as brightness and color, change due to various astrophysical processes. These variations, collectively referred to as transient events, encompass explosive phenomena (e.g., supernovae, kilonovae, gamma-ray bursts) and periodic stellar variability (e.g., flares, pulsating stars). The study of transient sources reveals the nature of extreme energy release and physical processes that are essential to understanding the formation and evolution of the universe. As a precursor to China's next-generation time-domain survey, the Mini-SiTian (MST) project aims to detect and classify these transient events \citep{he2024miniSiTian}.

The MST project uses a three-stage variable source detection process \citep{Gu2024}: first, the HOTPANTS image subtraction algorithm \citep{Becker2015HOTPANTS} is used to remove static background sources, generating a difference image. Next, candidate transient sources are extracted from the residual image using SExtractor \citep{bertin1996sextractor}. Finally, a manual screening process is employed to identify reliable samples. However, due to multiple interfering factors, such as optical artifacts, bright star spikes, mismatches during image subtraction, and instrumental effects, the image subtraction algorithm cannot perfectly eliminate static background sources. This leads to the generation of a large number of candidate sources during the source extraction phase. Observational data show that MST produces hundreds of thousands of transient alerts every night, of which approximately 99\% are false positives. This high false positive rate and inefficient manual screening process pose significant challenges to the scientific goals of the project. Therefore, it is essential to develop an efficient, accurate, and automated classifier to distinguish true astrophysical events (real) from false positives (bogus).

In recent years, automated and intelligent deep learning methods, particularly convolutional neural networks (CNNs), have become one of the primary techniques for addressing the classification of transient sources as real or bogus (\citealt{goldstein2015automated, cabrera2017deephits, duev2019ztf, hosenie2021meercrab, killestein2021goto, chen2023transientvit}). These methods analyze the image features of candidate sources to assess the authenticity of transient sources, ultimately assigning a score to each source on a scale from 0.0 (bogus) to 1.0 (real).

In this paper, we propose a CNN-based ensemble classifier for the MST project, aiming to address the challenge of high false positive rates and the need for efficient real-bogus classification. A high-quality dataset has been constructed, consisting of 177,696 bogus samples and 3,000 real samples, each represented as a 64 × 64 pixel residual image. We then conducted a rigorous evaluation of thirteen CNN architectures, spanning classical architectures (e.g., VGGNet \citep{simonyan2014vggnet}, ResNet \citep{he2016resnet}, DenseNet \citep{huang2017densenet} ), lightweight designs (e.g., EfficientNet \citep{tan2019efficientnet}, ShuffleNetV2 \citep{ma2018shufflenet}), and emerging paradigms (e.g., RepVGG \citep{ding2021repvgg}, UniRepLKNet \citep{ding2024unireplknet}). Experimental analysis revealed that ResNet exhibited outstanding specificity (99.70\%), EfficientNet achieved the highest recall (98.68\%), and DenseNet demonstrated a balanced performance with a recall of 94.71\% and specificity of 98.66\%.

To leverage the strengths of these models, we integrated ResNet18, EfficientNet\_B0, and DenseNet121 using a bootstrap aggregating (Bagging) \citep{breiman1996bagging} strategy combined with soft voting, resulting in a highly robust classifier with an AUC of 0.9961. This ensemble classifier achieved an optimal balance on the test set, attaining a recall of 95.37\% and a specificity of 99.25\%. Through dynamic threshold optimization, we identified 0.798 as the optimal threshold, at which both precision and recall reached 92.2\%. Validated on 5,000 operational samples, the classifier demonstrated strong performance in practical deployment, achieving a precision of 88.31\%, a recall of 91.89\%, and a specificity of 99.82\%. The system has now been stably deployed in the MST real-time processing pipeline, providing crucial technical support for achieving the scientific objectives of the MST project.

The organization of this paper is as follows: Section \ref{sec:data} presents a detailed description of the dataset. Section \ref{sec3:methods} outlines the experimental methods. Section \ref{sec:results} offers an in-depth analysis of the experimental results. Finally, Section \ref{sec:conclusions} provides a concise summary.

\section{DATA}\label{sec:data}
\subsection{Mini-SiTian}
The MST is the pathfinder of the SiTian project \citep{liu2021sitian}, which aims to provide a technology validation platform and scientific target exploration for the next generation of time-domain surveys. Its core equipment includes three 30\,cm Schmidt Complex Achromatic Telescopes (MST1, MST2, and MST3), each equipped with a ZWO ASI6200MM Pro CMOS camera (9,576 $\times$ 6,288 pixels, 3.76~$\mu$m pixel size) and SDSS-like filters ($i^{\prime}$, $g^{\prime}$, $r^{\prime}$) with a pixel scale of 0.862$^{\prime\prime}$/pixel.

From the start of pilot observations in November 2022 to July 2024, the MST has completed a total of 552 operational days, of which 344 were successfully acquired. Its observational targets cover 92 general sky fields, 2 comets, 45 transient sources from the Transient Name Server (TNS), and 33 LIGO gravitational wave alerts \citep{he2024miniSiTian}. These observations provide the basis for the construction of subsequent datasets.

\begin{figure}
    \centering
    \includegraphics[width=0.5\linewidth]{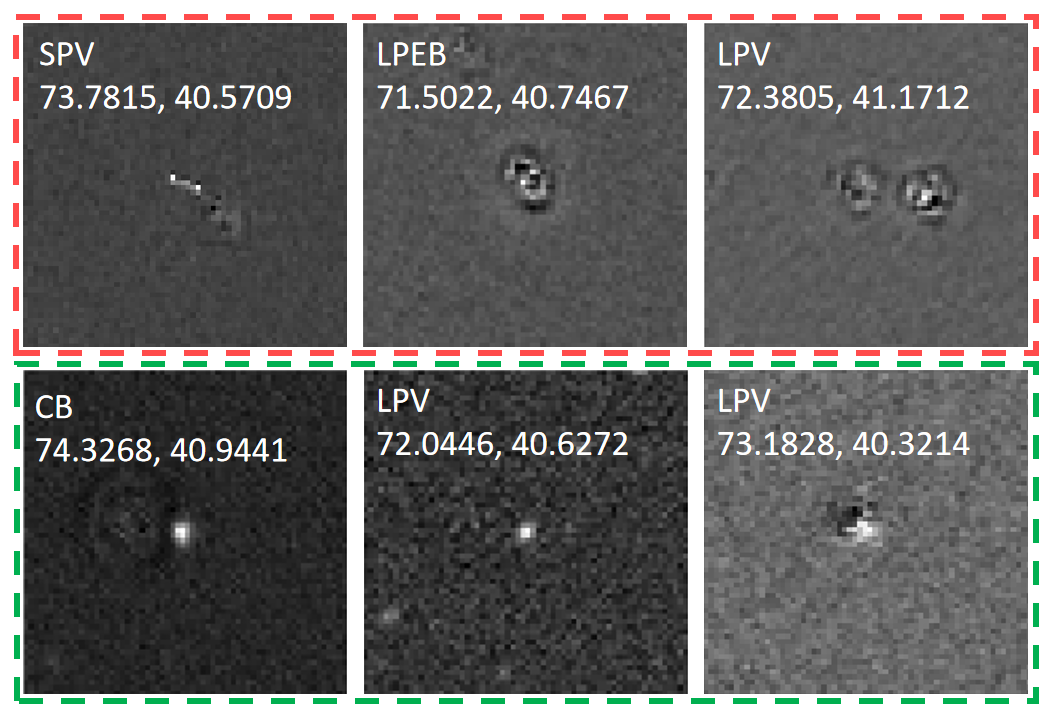}
    \caption{Examples of residual images of real candidates during the dataset filtering process include a Short Period Variable (SPV), a Long Period Eclipsing Binary (LPEB), a Contact Binary (CB), and three Long Period Variables (LPV). Each example is labeled with its type and its RA and Dec coordinates. The top row illustrates problematic images excluded during manual inspection. The bottom row displays high-quality real samples retained after screening.}
    \label{fig1:real}
\end{figure}

\begin{figure}[htbp]
    \centering
    \begin{subfigure}[b]{\textwidth} 
        \centering
        \includegraphics[width=0.49\linewidth]{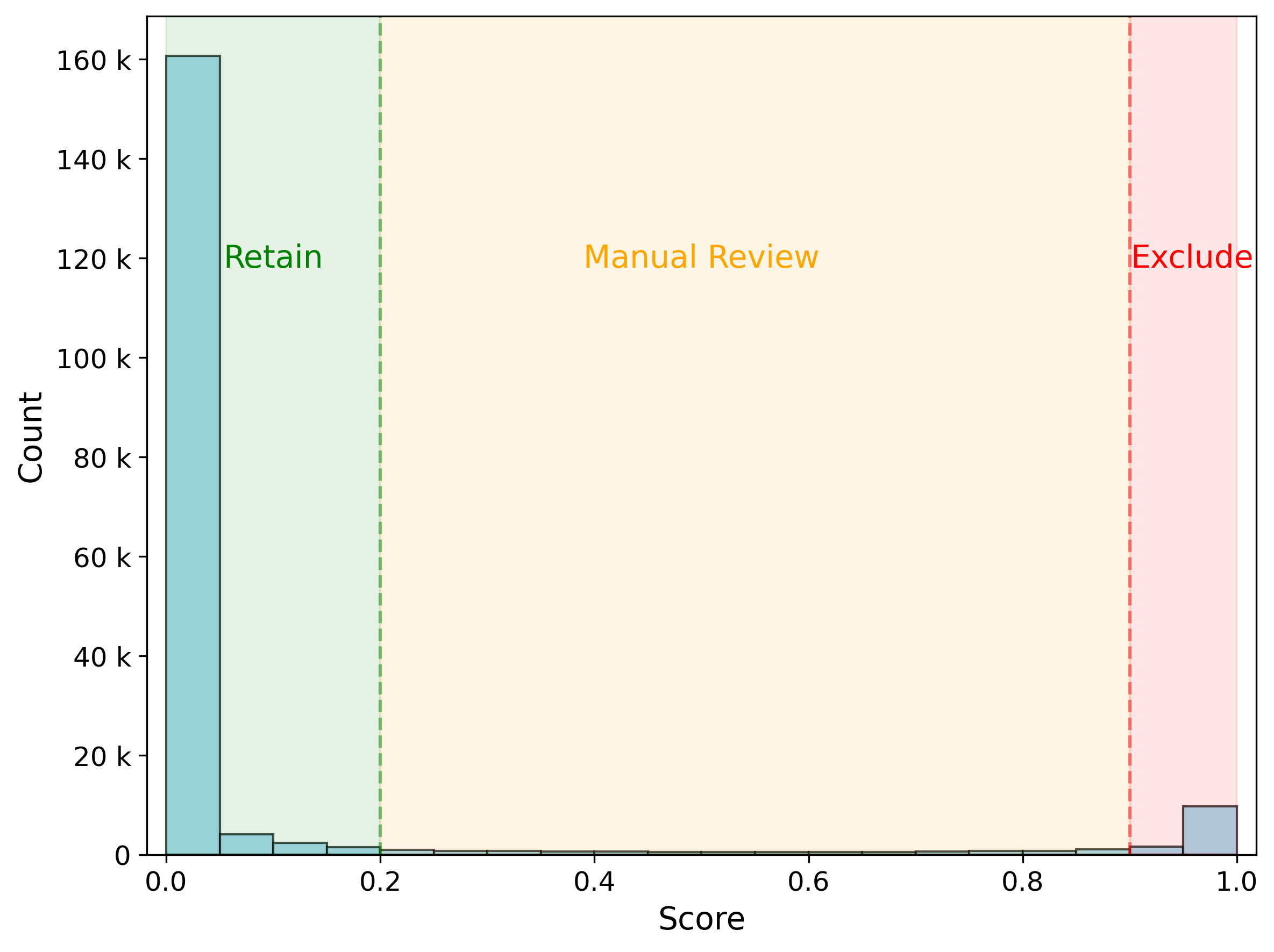}
        \caption{Distribution of classification scores for all bogus candidates, as predicted by the VGGNet13 classifier used for auxiliary selection. The x-axis represents the classification score, where higher values indicate a greater likelihood of the target being classified as real.}
        \label{fig2a:prob-distribution} 
    \end{subfigure}
    \vspace{0.5cm} 
    \begin{subfigure}[b]{\textwidth} 
        \centering
        \includegraphics[width=0.49\linewidth]{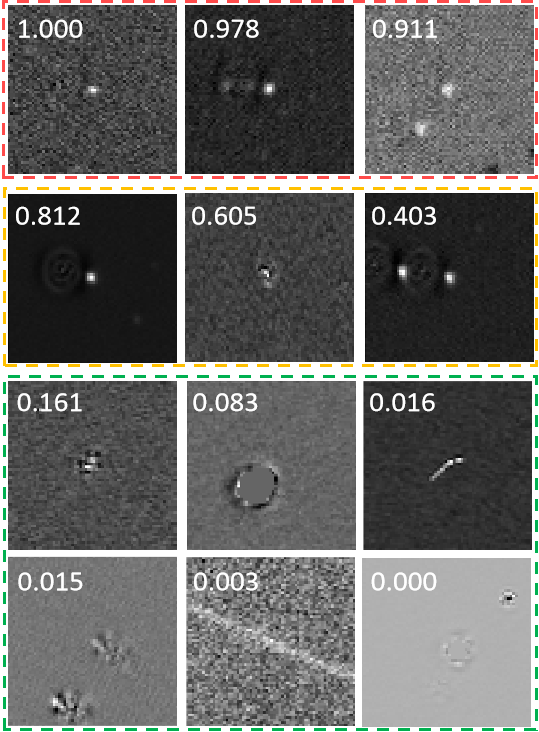}
        \caption{Examples of residual images of bogus candidates. The value in the upper left corner indicates the real-bogus score obtained using the auxiliary filtering classifier. The top row illustrates high-score samples ($>$0.9), which were excluded. The second row shows intermediate-score samples ($0.2–0.9$), which were manually reviewed. The bottom two rows show low-score samples ($<$0.2), which were directly retained.} 
        \label{fig2b:bogus-samples} 
    \end{subfigure}
    \caption{Distribution of classification-based scores and representative examples of bogus samples.} 
    \label{fig2:bogus} 
\end{figure}
\subsection{Dataset Creation}\label{sec:dataset_creation}
In supervised deep learning, model performance heavily relies on accurately labeled, reliable, consistently distributed, and representative samples. Based on observational data accumulated during the pilot phase of the MST, we designed a rigorous sample selection protocol that combines automated algorithms with manual review to construct a high-quality dataset.

\subsubsection{Real Samples}
The construction of the real sample set was based on continuous observations of 127 confirmed variable sources from the MST\citep{Gu2024}. To ensure data diversity, we randomly sampled observations from various epochs within each variable source's observation history. Using their coordinates, we extracted 64 × 64 pixel (55.16 arcsecond) cutout stamp images from the corresponding residual images to form the initial dataset. 

Despite these stamp images being derived from identified variable sources, there were some problematic samples. For example, some sample centers showed significant spurious features due to suboptimal image subtraction, while others lacked source features due to weak variability signals. To improve dataset reliability, we first used SExtractor to detect all stamp images, discarding those without visible central targets. Subsequently, we performed a visual review of the remaining samples and filtered out those with obvious false features.

As a result, a total of 3,000 real samples were obtained. Figure \ref{fig1:real} shows the problematic examples that were excluded and the high-quality examples that were retained.

\subsubsection{Bogus Samples}
To build the bogus sample dataset, we used SExtractor to automatically detect sources from over 200 residual images. To ensure data quality, we excluded detections near known variable sources and performed additional filtering on 190,000 samples.

An initial set of 3,000 bogus images was manually selected from the candidate dataset. These images, combined with the previously identified real samples, formed an initial training dataset. Using this dataset and the VGGNet13 network \citep{simonyan2014vggnet}, we got an auxiliary screening classifier, achieving an accuracy of 97\% on this initial dataset. As the training procedure aligns with that described in Section \ref{sec3:methods}, it is not detailed here. The classifier was then applied to all bogus candidates, yielding a probability score for each target to be classified as real.

Based on the probability scores provided by the classifier, we screened all candidate samples. Figure \ref{fig2a:prob-distribution} depicts the distribution of the predicted scores and the corresponding filtering strategies, which are detailed below:
\begin{enumerate}
    \item \textbf{High-score exclusion}: Targets with scores greater than 0.9 were excluded, as the model believed these samples should be classified as real, totaling 11,317. Visual inspection revealed that these targets were mainly residual signals from stationary objects that were not fully subtracted or alignment errors during image subtraction. They typically exhibited features nearly identical to the real samples, as shown in the first row of Figure \ref{fig2b:bogus-samples}.
    \item \textbf{Intermediate-score manual review}: Targets with scores between 0.1 and 0.8 were manually reviewed. Among the 9,994 inspected targets, 9,007 were retained. These targets showed intermediate confidence, indicating uncertainty in their classification as real or bogus, so they were carefully examined to ensure accurate labeling. The corresponding samples are shown in the second row of Figure \ref{fig2b:bogus-samples}.
    \item \textbf{Low-score retention}: Targets with classification scores below 0.2 were directly retained, as the model was highly confident that these sources were not real, and thus they were classified as bogus. A total of 168,689 samples were retained, which are shown in rows 3 and 4 of Figure \ref{fig2b:bogus-samples}.
\end{enumerate}

Finally, a total of 177,696 high-quality bogus samples were obtained.

\subsubsection{Dataset Composition}\label{sec:datasets}
Using the resulting real and bogus residual images, we constructed a dataset. The dataset was constructed for model training and validation, consisting of 177,696 bogus samples and 3,000 real samples, which were split into training, validation, and test sets with a ratio of 7:1:2.

\section{METHODS}\label{sec3:methods}
\subsection{CNNs and Ensemble Methods}
Convolutional neural networks (CNNs) are one of the most effective deep learning architectures for image-related tasks, including classification, object detection, and segmentation. By leveraging convolutional operations for feature extraction, pooling layers for dimensionality reduction, and fully connected layers for decision-making, CNNs enable high-precision image recognition. In recent years, various advanced CNN architectures such as VGGNet \citep{simonyan2014vggnet}, ResNet \citep{he2016resnet}, DenseNet \citep{huang2017densenet}, and EfficientNet \citep{tan2019efficientnet} have been proposed. These models achieve diverse balances between the number of parameters, computational efficiency, and performance, providing a wide range of choices for different application tasks.

To achieve accurate classification of real transient sources and bogus detections in the MST, we trained models with various architectures and parameters using the dataset introduced in Sec.\ref{sec:dataset_creation}. All models followed the training strategy described in Sec.\ref{sec:training_strategy}. The performance metrics for each model are detailed in Sec.\ref{sec:comparative-experiment}. Based on experimental results and analysis, we selected ResNet18, DenseNet121, and EfficientNet\_B0 as base learners and constructed a more powerful ensemble classifier using the bagging ensemble learning method.

Specifically, during training, bootstrap resampling was applied to the training data to generate multiple training subsets with the same sample size as the original training set. This resampling method allows certain samples to appear multiple times in the new subsets, while other samples may be excluded, thus introducing diversity into the training data. Given the class imbalance in the dataset, resampling is performed separately for each class. Each base model is independently trained on its corresponding resampled subset to ensure diversity and variability among the models \citep{breiman1996bagging}.

\begin{figure*}
    \centering
    
    \includegraphics[width=1\linewidth]{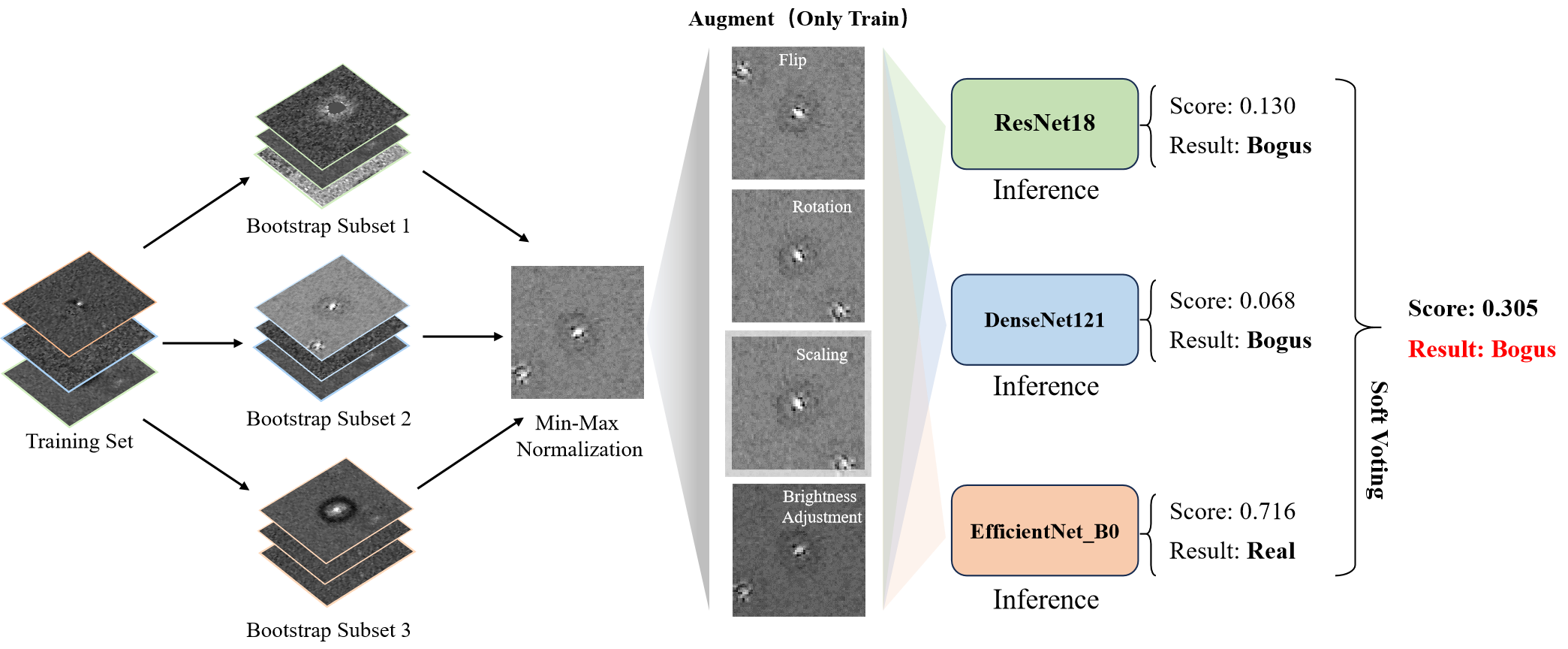}
    \caption{Overview of the ensemble classifier building process. The training set is divided into bootstrap subsets, and each subset is used to train ResNet18, DenseNet121, and EfficientNet\_B0 with data augmentation. During inference, predictions are combined using a soft voting strategy to achieve final classification results.}
    \label{fig3:model}
\end{figure*}

During the inference phase, we employed a soft voting strategy by averaging the predicted probabilities from all models to generate the final prediction. This approach effectively combines the strengths of multiple models while mitigating the weaknesses of individual models, significantly improving the classifier's performance. Figure \ref{fig3:model} provides an intuitive illustration of this method, including bootstrap resampling of training data, individual model training, and the soft voting strategy during the inference phase.
\subsection{Training Strategy}\label{sec:training_strategy}
In the training of CNNs, the selection of hyperparameters and the design of training strategies are critical to the performance and stability of the models \citep{mishkin2017systematic}. To ensure consistency and fairness in the training processes of the multiple base models used in this paper, we adopted a uniform hyperparameter configuration for them, as detailed in Table \ref{tab:hyperparameters}. Furthermore, to mitigate the issue of model overfitting, a dropout rate of 0.25 was applied to the convolutional layers and 0.5 to the fully connected layers across all models.

\begin{table}
\centering
\caption{Hyperparametric configurations during the training phase.}
\label{tab:hyperparameters}
\begin{tabular}{lc}
\hline
\textbf{Parameter}               & \textbf{Value}                              \\ \hline
Total Epochs                   & 100                                       \\
Batch Size                       & 256                                       \\
Optimizer                        & AdamW \citep{hutter2017adamw}                                     \\
Weight Decay                     & 0.001                                     \\
Beta Values (\(\beta_1, \beta_2\)) & (0.9, 0.999)                            \\
Scheduler          & CosineAnnealingLR                         \\
Initial Learning Rate (\(lr\))   & \(1 \times 10^{-4}\)                                     \\
Minimum Learning Rate            & \(1 \times 10^{-7}\)                      \\ 
\hline
\end{tabular}
\end{table}

\subsubsection{Data Augmentation}
To enhance model generalization and mitigate overfitting, we applied online data augmentation techniques during training. These techniques generate new samples by applying random transformations to existing ones, thereby increasing the diversity of the training dataset. In this paper, we selected the following augmentation methods based on the characteristics of astronomical images:
\begin{enumerate}
\item[a)] \textbf{Flip}: Images were flipped along vertical or horizontal axes to simulate different observational perspectives.
\item[b)] \textbf{Rotation}: Images were randomly rotated by multiples of 90° to account for orientation variations.
\item[c)] \textbf{Scaling}: Images were resized randomly between 80\% and 120\% of their original dimensions to simulate varying observational distances or resolutions.
\item[d)] \textbf{Brightness Adjustment}: The brightness of images was adjusted within a range of 80\% to 120\% to simulate varying exposure conditions.
\end{enumerate}
    
Each augmentation method was applied with a 50\% probability.

\subsubsection{Class Balancing}
Class imbalance, where one class significantly outnumbers another, poses a substantial challenge as it biases the model towards the majority class. In our dataset, the ratio of negative (bogus) samples to positive (real) samples was 59:1, an extreme imbalance. To address this issue, we first increased the representativeness of the minority class (real samples) by duplicating instances in the training set, expanding their presence to five times the original amount. This ensured that the model received adequate exposure to minority-class samples during training. To counter the risk of overfitting associated with direct repetition, we integrated data augmentation techniques to enhance the diversity of the oversampled samples.

Furthermore, to further address the issue of class imbalance, we employed the focal loss function \citep{he2017focalloss}, an improvement over the standard cross-entropy loss, which dynamically adjusts the model's focus on different samples. The focal loss is defined as:

   \begin{equation}
   FL(p_t) = -\alpha_t (1 - p_t)^\gamma \log(p_t)
   \end{equation}
 where \(p_t\) represents the predicted probability of the true class. \(\alpha_t\) is the class weighting factor, set to \([0.1, 0.9]\) in this paper to emphasize the minority class. \(\gamma\) is a focusing parameter, set to 2.0, which amplifies the loss contribution of hard-to-classify samples. By adjusting \(\alpha_t\) and \(\gamma\), focal loss reduces the contribution of well-classified samples, improving performance on the minority class. 
   
The combination of oversampling and focal loss significantly improved the model's ability to learn from imbalanced datasets.

\subsection{Experimental Details} \label{sec:exp}
The experimental setup consisted of a system with two Intel Xeon Gold 6230 processors, an NVIDIA A100-PCIE-40GB GPU running CUDA 12.2, and 251 GB of memory. All experiments were implemented in the Python programming language. 

\begin{figure*}[htbp]
    \centering
    \begin{subfigure}[b]{0.45\textwidth} 
        \centering
        \includegraphics[width=\linewidth]{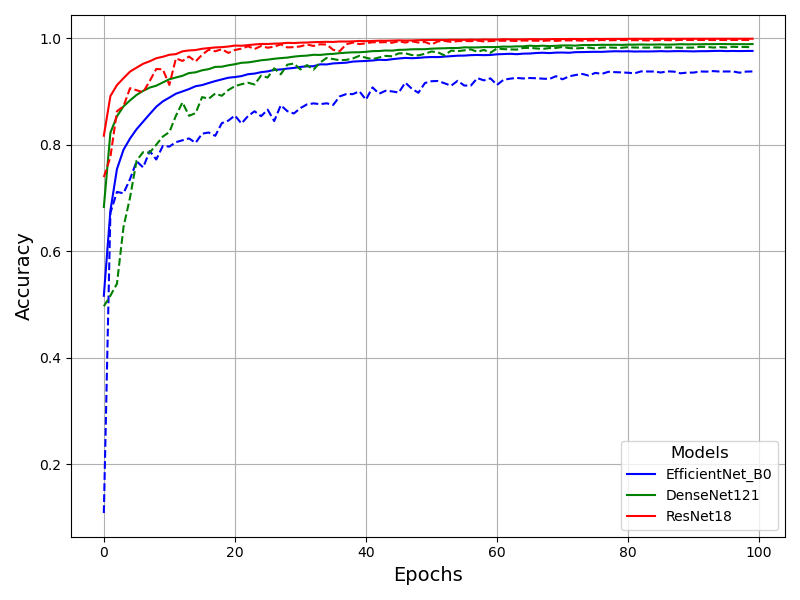}
        \caption{Accuracy over epochs} 
        \label{fig4a:accuracy} 
    \end{subfigure}
    \hfill 
    \begin{subfigure}[b]{0.45\textwidth} 
        \centering
        \includegraphics[width=\linewidth]{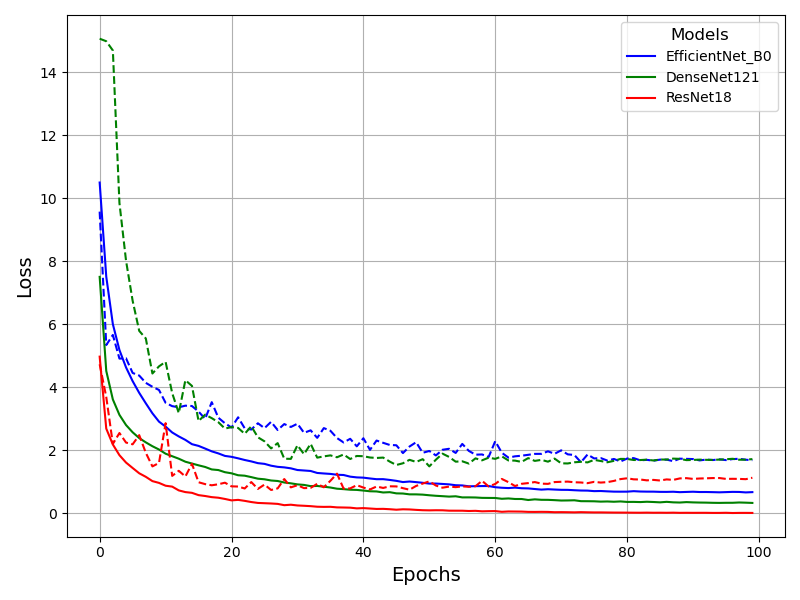}
        \caption{Loss over epochs}
        \label{fig4b:loss}
    \end{subfigure}
    \caption{Training and validation curves for ResNet18, DenseNet121, and EfficientNet\_B0 over 100 epochs of bagging ensemble training. Solid lines represent training metrics, and dashed lines represent validation metrics.}
    \label{fig4:acc-loss-cruve} 
\end{figure*}

Figure \ref{fig4:acc-loss-cruve} illustrates the evolution of loss and accuracy values for the three base models during the construction process of the ensemble classifier. As the epochs progress, a consistent decline in both training and validation loss can be observed, accompanied by a steady rise in accuracy until it converges. These results suggest that the models were effectively trained, exhibiting stable convergence behavior without signs of overfitting or underfitting throughout the process.

\section{RESULTS}\label{sec:results}
\subsection{Evaluation Metrics}
To evaluate the model's ability to distinguish between real and bogus samples, the following performance metrics were employed:  

\textbf{Specificity}: This metric assesses the model's ability to correctly identify bogus samples. A high specificity ensures that the model effectively reduces the misclassification of bogus as real, thereby avoiding unnecessary follow-up observations and resource consumption. It is mathematically expressed as:  
    \begin{equation}
        \text{Specificity} = \frac{\text{TN}}{\text{TN} + \text{FP}}
    \end{equation} 
where \(\text{TN}\) (True Negative) is the number of correctly classified bogus samples, and \(\text{FP}\) (False Positive) is the number of bogus misclassified as real. Similarly, \text{TP} (True Positive) refers to the number of correctly classified real and \text{FN} (False Negative) refers to the number of real misclassified as bogus.

\textbf{Recall}: This metric measures the model's ability to correctly identify positive samples. A high recall indicates that the model effectively minimizes the misclassification of true transients as bogus. This metric is particularly critical, as missing real samples may result in the loss of subsequent scientific targets. Its formula is as follows: 
    \begin{equation}
        \text{Recall} = \frac{\text{TP}}{\text{TP} + \text{FN}}
    \end{equation}

\textbf{Precision}: This metric quantifies the reliability of positive predictions. High precision ensures efficient allocation of observational resources.Its formula is as follows: 
    \begin{equation}
        \text{Precision} = \frac{\text{TP}}{\text{TP} + \text{FP}}
    \end{equation}

\textbf{F1-score}: The metric is the harmonic mean of precision and recall, evaluating the trade-off between minimizing missed transients and reducing bogus triggers. Its formula is as follows: 
    \begin{equation}
        \text{F1-score} = \frac{\text{2}\times\text{Recall}\times\text{{Precision}}}{\text{Recall}+\text{Precision}}
    \end{equation}

\textbf{Area Under Curve (AUC)}: The AUC serves as a quantitative metric of the model's overall classification performance. It is obtained by calculating the area under the ROC curve, which is plotted with the true positive rate (TPR) on the vertical axis and the false positive rate (FPR) on the horizontal axis at various classification thresholds. A value closer to 1 indicates better model performance. The formulas for TPR, FPR, and AUC are as follows:
\begin{equation}
\text{TPR} = \frac{\text{TP}}{\text{TP} + \text{FN}}
\end{equation}
\begin{equation}
\text{FPR} = \frac{\text{FP}}{\text{FP} + \text{TN}}
\end{equation} 
\begin{equation}
\text{AUC} = \int_{0}^{1} \text{TPR}(\text{FPR}) \, d\text{FPR}
\end{equation}

\begin{table*}
\setlength\tabcolsep{4pt}
\caption{Performance comparison of thirteen different CNNs on MST real-bogus classification. Models are sorted in descending order of AUC to highlight high-performing architectures.}
\centering
\begin{tabular}{lcccccc}
\hline
\textbf{Model}       & \textbf{Parameters (M)} & \textbf{Specificity (\%)} &\textbf{Recall (\%)} & \textbf{Precision (\%)} & \textbf{F1-score (\%)} & \textbf{AUC} \\
\hline
ResNet50            & 25.6           & 99.70            & 91.74       & 83.96          & 87.68         & 0.9948 \\      
ResNet18            & 11.7           & 99.70            & 91.40       & 83.92          & 87.50         & 0.9948 \\
ResNet34            & 21.7           & 99.69            & 91.40       & 83.41          & 87.22         & 0.9946 \\
EfficientNet\_B0    & 5.3            & 94.09            & 98.68       & 22.14          & 36.16         & 0.9932 \\
EfficientNet\_B2    & 9.2            & 94.64            & 98.18       & 23.78          & 38.29         & 0.9929 \\
DenseNet121         & 8.0            & 98.66            & 94.55       & 54.58          & 69.21         & 0.9928 \\
EfficientNet\_B1    & 7.8            & 93.98            & 98.68       & 21.82          & 35.74         & 0.9926 \\
DenseNet169         & 14.1           & 98.60            & 94.71       & 53.55          & 68.42         & 0.9926 \\
VGGNet11            & 138.0          & 97.46            & 92.07       & 38.18          & 53.97         & 0.9903 \\
UniRepLKNet\_A      & 4.4            & 98.36            & 90.25       & 48.40          & 63.01         & 0.9883 \\
ViT\_Tiny           & 5.7            & 91.45            & 94.05       & 15.77          & 27.02         & 0.9736 \\
ShuffleNetV2\_x0\_5 & 1.4            & 99.64            & 12.89       & 38.05          & 19.26         & 0.7597 \\
RepVGG\_A0          & 8.3            & 94.05            & 26.12       & 6.95           & 10.98         & 0.5616 \\
\hline
\end{tabular}
\label{tab:single-model-metrics}
\end{table*}

\subsection{Classifier Performance}\label{sec:classifier}
\subsubsection{Individual CNN Classifier}\label{sec:comparative-experiment}

Table \ref{tab:single-model-metrics} summarizes thirteen CNN performance metrics on the test set under a threshold of 0.5, sorted by AUC value from largest to smallest. From the results, the ResNet series emerged as top performers in comprehensive metrics, achieving leading AUC values (0.9946–0.9948) and F1-scores (87.22\%–87.68\%). Notably, their specificity reaches 99.70\%, demonstrating exceptional capability in bogus sample identification. However, their recall (91.40\%-91.74\%) still needs to be improved for MST real-bogus classification scenarios. The EfficientNet series attained the highest recall rates (98.18\%–98.68\%) across all models. Nevertheless, extreme predictive bias toward real classes severely constrained their precision (21.82\%–23.87\%) and F1-scores (35.74\%–38.29\%). In contrast, the DenseNet series achieved a relatively balanced trade-off between recall (94.55\%–94.71\%) and specificity (98.60\%–98.66\%), but its overall performance still leaves room for improvement. The performance of the remaining models was generally inferior to the aforementioned three models. Particularly, ShuffleNetV2 and RepVGG exhibited suboptimal performance across multiple metrics, which suggests that their architectures may not be well-suited for the current task.

Furthermore, the experimental results indicate that there is no significant correlation between model performance and parameter count. Despite possessing 138M parameters, making it the model with the largest parameter set, VGGNet11 underperformed across multiple metrics compared to models with fewer parameters, such as ResNet. Moreover, different parameter versions within the same model family (e.g., ResNet18 and ResNet34, EfficientNet\_B0 and EfficientNet\_B2) exhibited comparable performance on key metrics, further highlighting the redundancy of model parameters. This suggests that, in the context of this task, selecting models with varying parameter scales within the same family has a limited impact on performance. Conversely, lower-parameter models offer advantages in terms of deployment and operational efficiency, presenting a more optimal choice for practical applications.

\subsubsection{Ensemble Classifier}
\begin{table*}
\setlength\tabcolsep{4pt}
\caption{Performance evaluation of bagging ensembles for MST real-bogus classification, covering individual bagged models and combinations of ResNet18, DenseNet121, and EfficientNet\_B0.}
\centering
\begin{tabular}{lcccccc}
\hline
\textbf{Model} & \textbf{Specificity (\%)} & \textbf{Recall (\%)} & \textbf{Precision (\%)} & \textbf{F1-score (\%)} & \textbf{AUC} \\
\hline
ResNet18                                & 99.85            & 91.90       & 91.00          & 91.45         & 0.9951          \\
EfficientNet\_B0                        & 94.00            & 98.35       & 21.82          & 35.71         & 0.9928          \\
DenseNet121                             & 98.76            & 94.38       & 56.48          & 70.67         & 0.9926          \\
DenseNet121, EfficientNet\_B0           & 96.99            & 96.03       & 35.21          & 51.53         & 0.9938          \\
EfficientNet\_B0, ResNet18              & 98.33            & 95.04       & 49.23          & 64.86         & 0.9953          \\
DenseNet121, ResNet18                   & 99.30            & 94.54       & 69.76          & 80.28         & 0.9955          \\
DenseNet121, EfficientNet\_B0, ResNet18 & 99.25            & 95.37       & 68.45          & 79.70         & \textbf{0.9961} \\
\hline
\end{tabular}
\label{tab:ensemble-metrics}
\end{table*}
To further develop a more robust classifier that meets the requirements of the MST project, we conducted bagging training with three models - ResNet18, EfficientNet\_B0, and DenseNet121 - based on the experimental analysis presented in section \ref{sec:comparative-experiment}. Subsequently, we implemented ensemble learning through soft voting strategies with different model combinations and evaluated their performance metrics. Table \ref{tab:ensemble-metrics} comprehensively documents the performance of individual retrained models and the ensemble results of various model combinations.
\begin{figure}[t]
    \centering
    \includegraphics[width=0.7\linewidth]{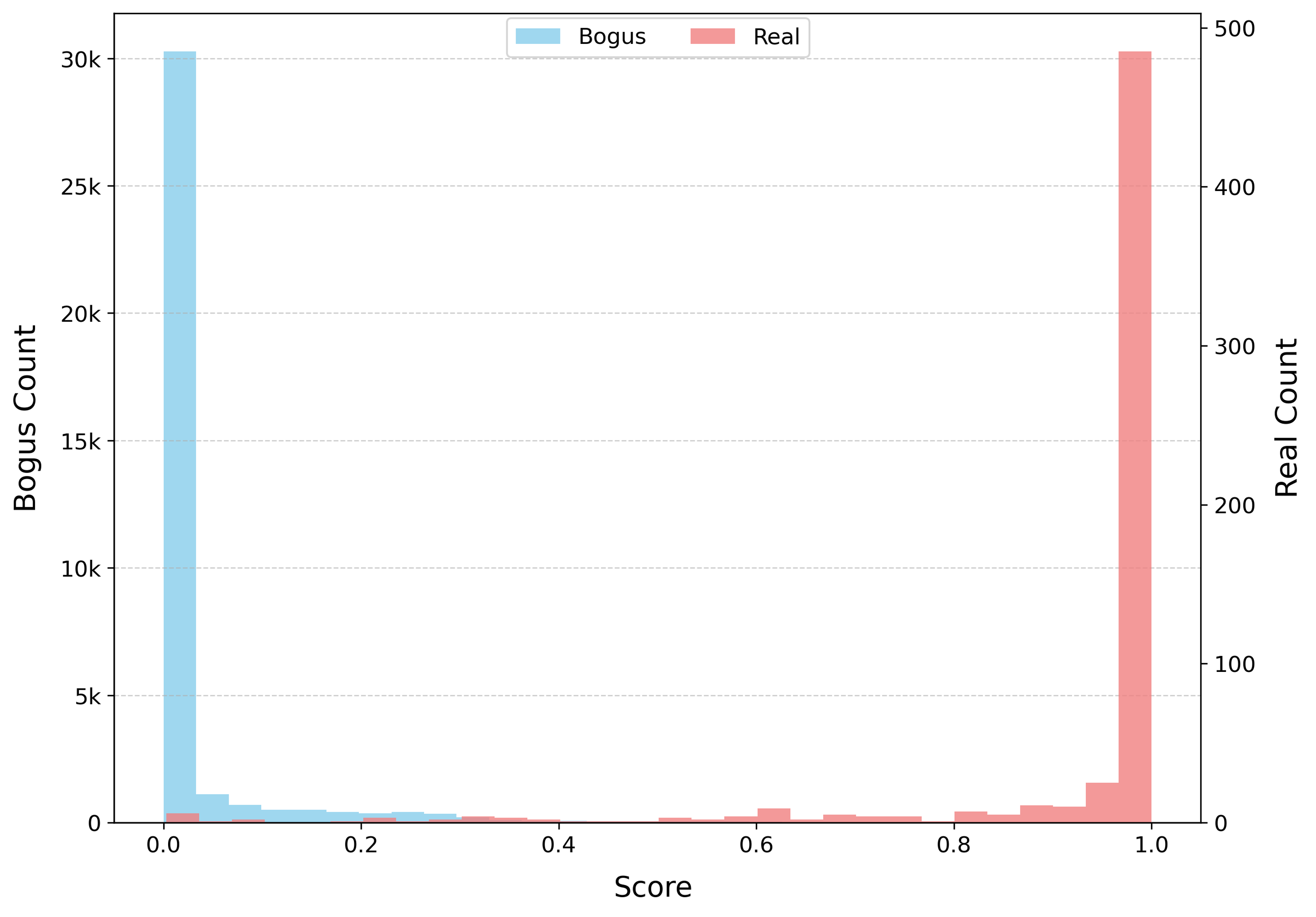}
    \caption{Histogram of ensemble model scores for 605 real samples and 35,538 bogus samples from the test set. The horizontal axis represents the prediction score, the left side of the vertical axis represents the number of bogus samples, and the right side represents the number of real samples.}
    \label{fig5:test-prob-dist}
\end{figure}

The experimental results demonstrate that the retrained models maintain strong consistency with previous training outcomes (record in table \ref{tab:single-model-metrics}), thereby validating the effectiveness of ensemble training. Notably, the tri-model ensemble configuration comprising DenseNet121, EfficientNet\_B0, and ResNet18 achieved the highest AUC value of 0.9961. This finding indicates that this particular ensemble exhibits superior overall classification performance compared to all tested model combinations. Although the ensemble model shows slight decreases in F1-score (79.70\%) and precision (68.45\%) metrics relative to some other combinations, the significant improvement in AUC, coupled with maintained excellence in recall (95.37\%) and specificity (99.25\%), substantiates its enhanced discriminative power between real and bogus samples. Given that integrating heterogeneous architectures effectively enhances model generalization capabilities and robustness while ultimately achieving optimal overall performance, we ultimately selected the ensemble configuration of DenseNet121, EfficientNet\_B0, and ResNet18 as our preferred solution.
\begin{figure}[t]
    \centering
    \includegraphics[width=0.7\linewidth]{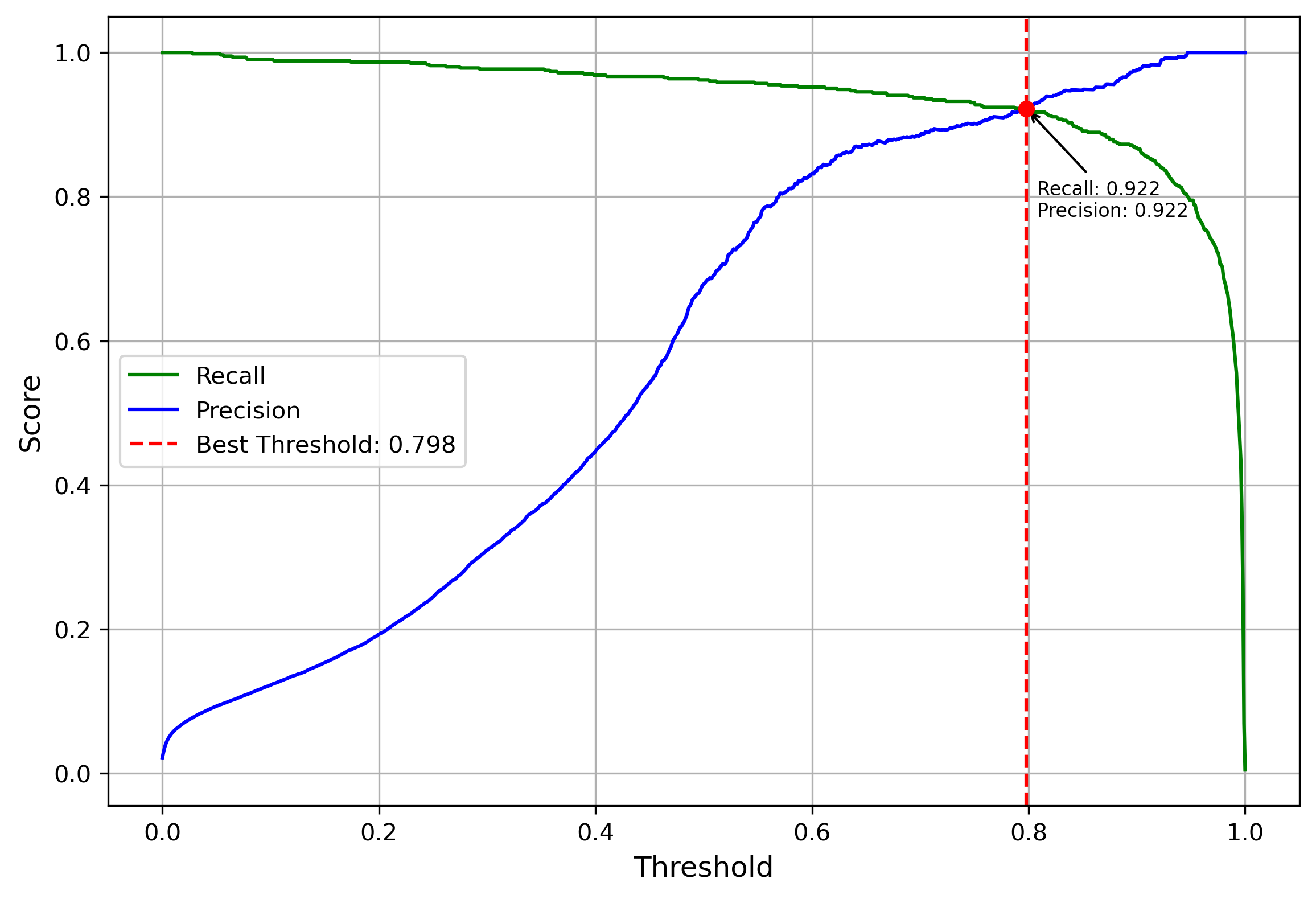}
    \caption{Precision and recall curves of the ensemble model as a function of the classification threshold. The vertical dashed red line highlights the optimal threshold (0.798), where precision and recall both reach 0.922, indicating a balanced performance.}
    \label{fig6:recall-precision}
\end{figure}

Figure \ref{fig5:test-prob-dist} presents a bimodal histogram comparing the predicted score distributions of the ensemble model for the real (605 samples) and bogus (35,538 samples) classes in the test set. As depicted in the figure, the distributions of the real (red) and bogus (blue) samples exhibit a clear separation. The bogus samples form a sharp peak near 0, with over 95\% concentrated within the 0-0.2 score range. Conversely, the real samples demonstrate a dominant peak near 1, with the majority of predictions clustering between 0.8 and 1.0. Only a minority of samples reside within the intermediate probability range of 0.2 to 0.8. This evident separation suggests the classifier is capable of making high-confidence classification decisions in most instances. However, the small number of samples in the intermediate range reflects the model's uncertainty in handling boundary cases, suggesting potential directions for further optimization.

In binary classification tasks, the classification threshold is a critical parameter that determines how the model assigns samples to the positive or negative class. Figure \ref{fig6:recall-precision} visualizes the variations in precision and recall of the ensemble model under different thresholds, thereby assisting in the selection of the optimal decision point. As illustrated in the figure, the ensemble model achieves its best balanced performance at a threshold of 0.798, where both precision and recall reach an impressive 0.922. This indicates that by setting the threshold to 0.798, the model can effectively identify real samples with high precision while minimizing the risk of missing true positives. This not only further solidifies the robust capability of the ensemble model in accurately distinguishing between real and bogus samples but also significantly enhances its reliability in practical application scenarios. In the actual deployment of the MST project, 0.798 serves as a crucial reference for the initial threshold, which is dynamically adjusted around this value based on specific task requirements.

\subsection{Real-World Application}

\begin{figure}[t]
    \centering
    \includegraphics[width=0.4\linewidth]{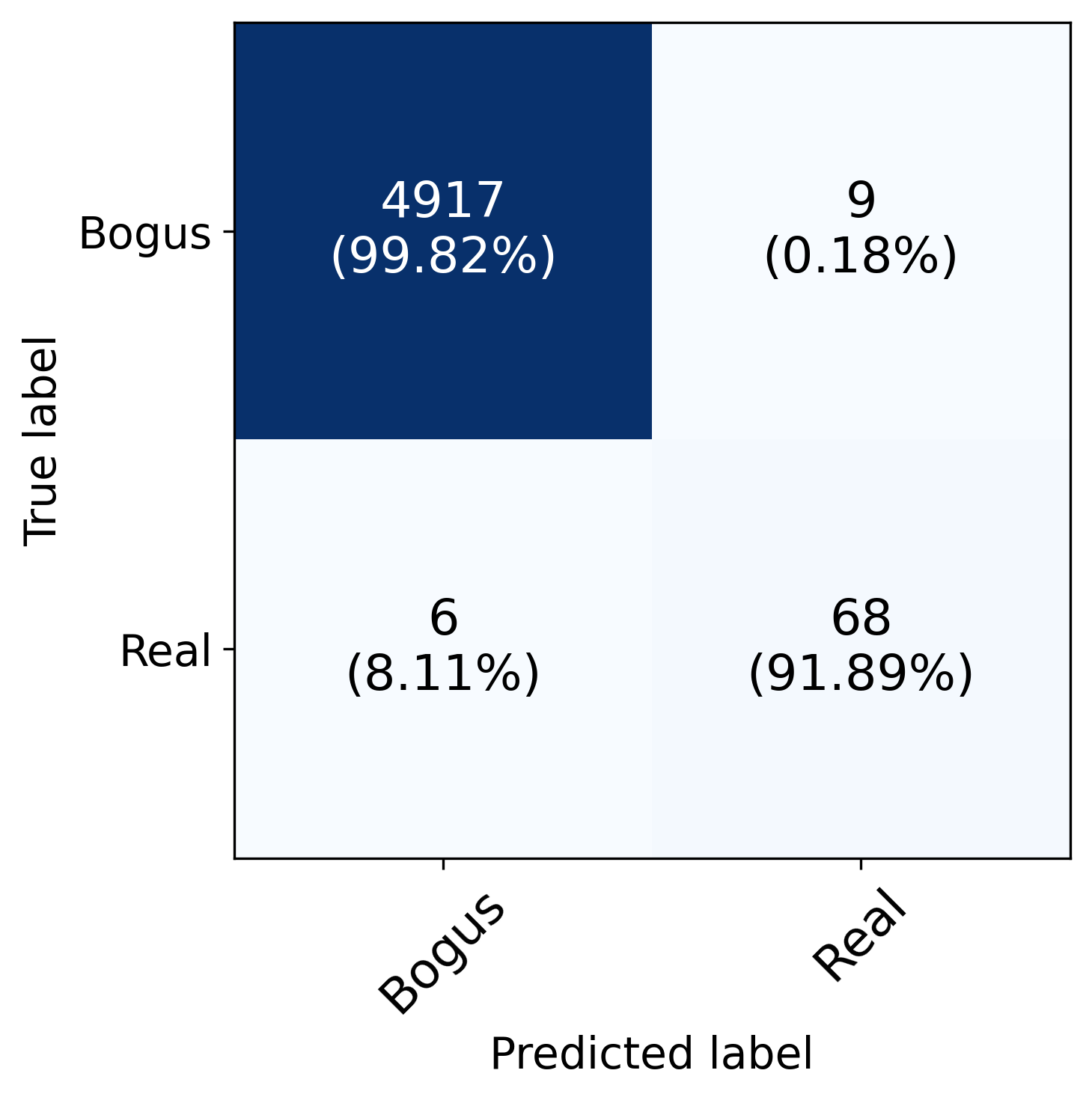}
    \caption{Confusion matrix of the ensemble model on 5,000 randomly selected real-world application samples. The ground truth is determined through manual inspection, while the predicted labels are generated by the classifier.}
    \label{fig7:real-world-cm}
\end{figure}
The ensemble classifier proposed in this study has been successfully deployed in the MST real-time data processing pipeline. Based on previous experimental analyses, we set the classification threshold at 0.798 to balance precision and recall in practical applications. To further validate the operational effectiveness of the classifier, we randomly selected 5,000 samples from actual operations and conducted manual verification of the classification results. Among these samples, the classifier identified 77 as real and 4923 as bogus. Using the manually reviewed results as the ground truth, we obtained the confusion matrix shown in Figure \ref{fig7:real-world-cm}, which reflects the model's classification performance in real deployment environments.

Based on calculations derived from this confusion matrix, the model achieved a precision of 88.31\%, a recall of 91.89\%, and a specificity of 99.82\% during practical use. These results indicate a slight degradation in performance relative to the optimal values observed during the validation phase. This discrepancy can be attributed to subtle differences between the distributions of the operational and training data, as well as variations in the proportion of genuine to bogus samples encountered in real-world scenarios. Nonetheless, the model maintains a high level of performance, effectively detecting true targets while mitigating the risk of false alarms.

In conclusion, the model’s performance in real-world applications is sufficient to meet the operational requirements of the MST project, thereby providing robust technical support for its efficient and stable implementation. As the classifier continues to operate within the MST project, we will progressively accumulate more real-world operational data. These valuable datasets will be systematically incorporated to enhance and optimize our training corpus, thereby driving iterative improvements in model performance through continuous learning cycles.

\section{CONCLUSIONS}\label{sec:conclusions}
In this study, we developed a deep learning-based ensemble classifier specifically designed to distinguish real transient sources from bogus detections in the MST. To support this work, we constructed a dataset. The first dataset, used for model training and evaluation, comprised 177,696 bogus samples and 3,000 real samples, carefully selected through a combination of automated filtering and manual inspection.

Furthermore, we conducted a comprehensive evaluation of 13 CNN models and, based on the experimental results, adopted a bagging strategy to organically integrate ResNet18, EfficientNet\_B0, and DenseNet121 into an ensemble classifier. This classifier not only achieves an excellent AUC (0.9961) but also exhibits balanced performance, with a recall of 95.37\% and a specificity of 99.25\%. The prediction scores show a distinctly bimodal distribution: the vast majority of bogus samples cluster near 0, whereas real samples concentrate near 1, thereby effectively reducing decision uncertainty at the boundaries.

In addition, the classifier’s performance in real-world environments further validates its practical applicability. Although there is a slight degradation when applied to actual MST observational data, the model still attains an precision of 88.31\%, a recall of 91.89\%, and a specificity of 99.82\%. This outstanding performance robustly demonstrates the benefits of employing an ensemble of heterogeneous architectures.

Overall, the deep learning-based ensemble classifier we developed effectively meets the operational requirements of the MST project. Looking ahead, as more data are incorporated and the model undergoes continuous iterative optimization, its performance and adaptability are expected to improve further. These ongoing enhancements are poised to consolidate its critical role in the MST real-time processing pipeline and ultimately advance in-depth scientific research into dynamic astrophysical phenomena.

\normalem
\begin{acknowledgements}
This work is supported by the National Key Basic R\&D Program of China via 2023YFA1608303, the Strategic Priority Research Program of the Chinese Academy of Sciences (XDB0550103), and the National Natural Science Foundation of China under grants Nos. 12273076, 12133001, 12422303, and 12261141690.

The SiTian project is a next-generation, large-scale time-domain survey designed to build an array of over 60 optical telescopes, primarily located at observatory sites in China. This array will enable single-exposure observations of the entire northern hemisphere night sky with a cadence of only 30-minute, capturing true color (gri) time-series data down to about 21 mag. This project is proposed and led by the National Astronomical Observatories, Chinese Academy of Sciences (NAOC). As the pathfinder for the SiTian project, the Mini-SiTian project utilizes an array of three 30 cm telescopes to simulate a single node of the full SiTian array. The Mini-SiTian has begun its survey since November 2022. The SiTian and Mini-SiTian have been supported from the Strategic Pioneer Program of the Astronomy Large-Scale Scientific Facility, Chinese Academy of Sciences and the Science and Education Integration Funding of University of Chinese Academy of Sciences.

\end{acknowledgements}
  
\bibliographystyle{raa}
\bibliography{bibtex}

\end{document}